# Cryogenic characterization and modeling of a CMOS floating-gate device for quantum control hardware.

Michele Castriotta[1], Enrico Prati[2], Giorgio Ferrari[1*]

[1] (Dipartimento di Elettronica, Informazione e Bioingegneria, Politecnico di Milano, 20133 Milano, Italy)
[2] (Istituto di Fotonica e Nanotecnologie, Consiglio Nazionale delle Ricerche, Italy)
* Corresponding author. E-mail address: giorgio.ferrari@polimi.it

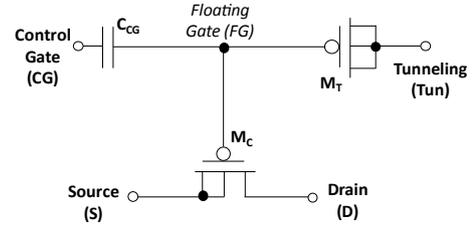

**Fig. 1** *Schematic of a p-type floating-gate device realized in standard CMOS technology. Both p-type transistors, $M_T$ and $M_C$, have a width of 400nm and a length of 350nm. The control gate capacitor $C_{CG}$ has a capacitance value of 50fF.*

***Abstract*** - We perform the characterization and modeling of a floating-gate device realized with a commercial 350-nm CMOS technology at cryogenic temperature. The programmability of the device offers a solution in the realization of a precise and flexible cryogenic system for qubits control in large-scale quantum computers. The device stores onto a floating-gate node a non-volatile charge, which can be bidirectionally modified by Fowler-Nordheim tunneling and impact-ionized hot-electron injection. These two injection mechanisms are characterized and modeled in compact equations both at 300 K and 15 K. At cryogenic temperature, we show a fine-tuning of the stored charge compatible with the operation of a precise analog memory. Moreover, we developed accurate simulation models of the proposed floating-gate device that set the stage for designing a programmable analog circuit with better performances and accuracy at a few Kelvin. This work offers a solution in the design of configurable analog electronics to be employed for accurately read out the qubit state at deep-cryogenic temperature.

*1. Introduction*: Quantum computing based on silicon [1, 2, 3] or superconductive qubits [4] operating at cryogenic temperature require silicon integrated circuits for controlling gate operations [5, 6]. A cryo-CMOS controller, operating at a few Kelvin, is a possible solution to address the interconnection complexity, cost, and poor reliability of a room-temperature control system [7, 8]. The cryogenic system is asked to read and generate both accurate and extremely low noise signals with a resolution better than 0.1mV. Such requirements mitigate perturbations of the qubit's fragile quantum state. A challenge of the cryo-CMOS design is to achieve these demanding specifications without impinging on the increased device mismatch at very low temperatures [9, 10]. Analog techniques, commonly used at room temperature to reduce the inaccuracy given by the mismatch, are auto-zeroing and chopper stabilization [11]. However, the resulting increase in power consumption and complexity could be incompatible with a cryogenic environment. A mismatch compensation based on a programmable element such as a floating-gate device is advantageous in compactness, minimal additional power, fine programmability, long-term retention, and CMOS compatibility. At room temperature, floating-gate devices are used as trimming elements to compensate for the amplifier offset by accurately adjusting the currents of the input differential stage [12]. Furthermore, an automatic offset compensation based on floating gate transistors allows the design of fast, accurate, and low-power comparators [13].

In standard CMOS technology, a floating gate is achievable by connecting the gate of a MOS transistor with a control gate capacitor, $C_{CG}$, without any change of the technological process, as shown in Figure 1. The charge onto the floating gate node, although electrically isolated from the surrounding, can be modified at room temperature using electron Fowler-Nordheim (FN) tunneling [14] and impact-ionized hot-electron injection (IHEI) [15]. Indeed, the p-type floating-gate device shown in Figure 1 is a four-terminal device comprising two p-type MOS transistors. The transistor $M_T$ allows the electron tunneling by applying a properly high voltage at the tunneling terminal (Tun). The transistor $M_C$ is used both for IHEI and readout of the stored charge by a charge-to-current conversion.

CMOS floating gate devices have been extensively characterized and investigated at room temperature [12, 13, 16]. However, at cryogenic temperature, CMOS floating gate devices have been studied only recently using large test structures (about 5 pF) in order to characterize the FN tunneling current and the IHEI current [17].
Here we demonstrate the cryogenic operation of a compact CMOS floating-gate device with a small gate capacitance (50 fF), and report its electrical characterization. The charge injection mechanisms are studied at 15 K and 300 K to understand the advantages these devices offer in the design of accurate cryogenic electronics. We show how the threshold voltage of such devices can be finely tuned by applying pulsed voltages to the cell, validating, at the same time, compact semi-empirical models for the FN tunneling and IHEI.

*2. Charge injection mechanisms*: FN tunneling and IHEI allow bidirectional memory updates, decreasing or increasing the charge accumulated onto the floating gate. We use FN tunneling through the oxide gate of $M_T$ to extract electrons from the floating node. A positive voltage is applied to the Tun terminal to create an electric field across the thin oxide of $M_T$ that increases the quantum transparency of the potential barrier. When this electric field is large enough, electrons tunnel off the floating gate, as shown in Figure 2b.

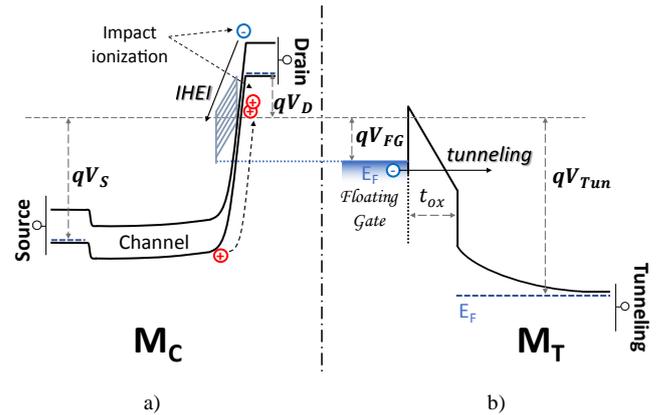

**Fig. 2** *a) Channel energy band diagram of the p-type MOS transistor $M_C$ for IHEI in the floating-gate. The energy band diagram from the drain to the FG is orthogonal to the source-drain diagram. b) Energy band diagram of the p-type MOS capacitor $M_T$ for the tunneling electrons off the floating gate. The energy level of the floating gate is the same in a and b.*

The tunneling current depends on the oxide voltage drop, according to the Fowler-Nordheim theory of field emission, as shown in the following equation [18]:

$$I_{Tun} = I_{Tun0} WL \left(\frac{V_{Tun} - V_{FG}}{t_{ox}}\right)^2 e^{-B \frac{t_{ox}}{V_{Tun} - V_{FG}}} \quad (1)$$

where $t_{ox}$ is the oxide thickness of $M_C$, $WL$ is its area, $V_{Tun}$ is the voltage applied to the terminal Tun and $V_{FG}$ the voltage of the floating-gate node. In Eq. (1) $I_{Tun0}$ and $B$ are two global fitting parameters dependent on temperature.

Although the tunneling mechanism could also be used to inject electrons in the floating node, in order to avoid a large negative voltage at the n-well of the $M_T$ transistor, we use IHEI to increase the number of electrons stored on the FG. The $M_C$ transistor is biased above threshold with a large drain-source voltage $V_{SD}$. The holes in the $M_C$ channel are accelerated from source to drain by the channel electric field. If the electric field in the depletion region between the channel and the drain is large enough, a fraction of these holes can collide with sufficient kinetic energy to break a Si-Si bond and create an additional electron-hole pair, as schematically shown in Figure 2a. The ionized electrons with kinetic energy higher than the gate oxide's energy barrier (∼3.1 $eV$), if scattered toward the gate oxide, can be injected in the floating-gate node. Based on the "lucky" electron model [19], successfully applied to the modeling of hot-electron injection in n-type MOS, the source of hot electrons in a p-type MOS is from the impact ionization process, which makes the gate current proportional to the drain current, which can be easily measured. The resulting equation describing the IHEI current [15, 20] depends on the maximum lateral channel electric field at the drain side, $E_m$, which can be expressed in terms of $V_{SD}$ as [15]:

$$E_m \approx \frac{V_{SD} - V_{Dsat}}{l}$$

where $l$ is the length where the hot-electron injection is significant and $V_{Dsat}$ is the saturation voltage due to the velocity saturation of the carriers and the pinch-off of the channel. To keep the equation simple, we substitute $V_{Dsat}$ with an equivalent overdrive voltage, $\delta \cdot V_{ov}$, where $\delta$ is a fitting parameter less than one because of the saturation velocity effect.
The resulting equation of the IHEI is given by:

$$I_{IHEI} = C(V_{SD} - \delta V_{ov})^3 I_{Drain} e^{-\frac{D}{(V_{SD} - \delta V_{ov})}} \quad (2)$$

where $V_{ov}=V_{FG}-V_S-V_{Th,0}$, $V_S$, $V_{Th,0}$ and $I_{Drain}$ are the overdrive voltage, the source voltage, the threshold voltage, and the drain current, respectively, of transistor $M_C$. $C$, $D$, and $\delta$ are global fitting parameters dependent on temperature. Eq. (2) models the IHEI only in terms of device terminal voltages in a suitable format for circuit simulation.

*3. Characterization methodology:* Room temperature (RT) and cryogenic measurements were performed on the p-type floating-gate device shown in Figure 3, realized in standard 350-nm CMOS technology by AMS. As many standard CMOS processes, this technology has all the n-type MOSFETs directly in the substrate, preventing the control of the body voltage of a single transistor. Therefore, a p-type MOSFET in an independent n-well was chosen as $M_T$ transistor, and the tunneling current was activated by applying a high voltage to the well. For the injection of hot electrons in the floating gate, a p-MOS transistor was preferred to a n-MOS for the higher injection efficiency at room temperature. This happens because for a p-MOS the oxide electric field at the drain side aids the electron injection when $|V_{GS}| \ll |V_{DS}|$ [20], a mechanism that should also be valid at low temperature.

We investigate the device operation in the low charge injection regime by monitoring the threshold voltage $V_{Th}$ of the device after each programming/erase phase, in which a small amount of charge is added or removed onto the floating gate node. This methodology allows to measure indirectly very low injection currents (10÷100 e⁻/ms), otherwise not measurable directly. After applying the voltages required to activate FN tunneling or IHEI for a time $\Delta t$, the device's threshold voltage was measured by controlling the transistor $M_T$ with the control gate terminal. The measured $V_{Th}$ value includes the threshold voltage $V_{Th,0}$ of the transistor $M_C$ and the effect of the charge stored in the floating gate. The floating-gate voltage $V_{FG}$ required to calculate the injection currents (Eqs. (1) and (2)) was obtained from the measured $V_{Th}$ as follows. The variation of the floating gate voltage at the n-th programming/erase phase, $\Delta V_{FG}(n)$, is given by:

$$\Delta V_{FG}(n) = \frac{C_{CG}}{C_T} \Delta V_{Th}(n) \quad (3)$$

where $\Delta V_{Th}(n)$ is the variation of the threshold voltage produced by the n-th application of the voltages for the programming/erase phase. $C_T$ is the total capacitance of the floating gate, which is equal to:

$$C_T = C_{CG} + C_{Tun} + C_S + C_D \quad (4)$$

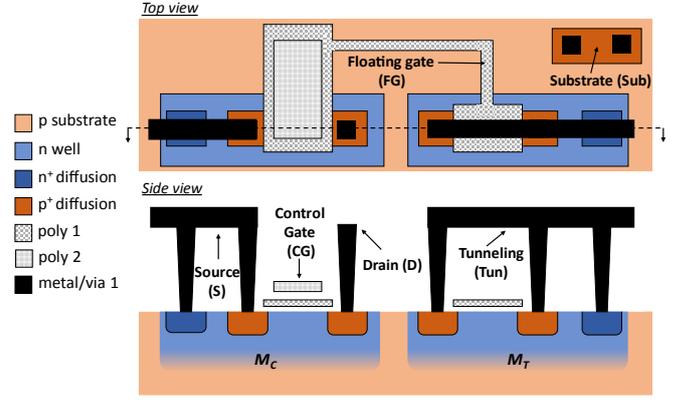

**Fig. 3** *Simplified layout of the p-type floating-gate device under test. The capacitive coupling to the floating-gate node is realized with the poly 2 control gate.*

$C_{CG}$ is the physical capacitance reported in Figure 1, and $C_{Tun}$, $C_S$ and $C_D$ are the parasitic capacitances between the floating gate node and the tunneling, source, and drain terminals, respectively. The value of $C_T$ was estimated at 64 fF using a physical process verification tool from the layout of the device under test to extract the parasitic capacitances.
The voltage on the floating gate is determined by two contributes: the capacitive coupling $V_{FG,CAP}$ given by the voltages applied to the device terminals, and the trapped charge $Q$ on the floating gate. Mathematically, we have:

$$V_{FG} = V_{FG,CAP} + \frac{Q}{C_T} \quad (5)$$

Deriving the Eq. (5) with respect to time, we can calculate the variation of the floating gate voltage, $V_{FG}$, during the programming/erase phase, obtaining the following differential equation:

$$\frac{dV_{FG}}{dt} = \frac{1}{C_T}\frac{dQ}{dt} = \frac{I_{TUN/IHEI}(V_{FG})}{C_T} \quad (6)$$

The solution of the Cauchy problem associated to Eq. (6) requires the assessment of the initial value of $V_{FG}$. By looking the Eq. (5), it is determined by the capacitive coupling and by the initial charge stored. The first contribution $V_{FG,CAP}$ is given by:

$$V_{FG,CAP} = V_{Tun,p}\frac{C_{Tun}}{C_T} + V_{S,p}\frac{C_S}{C_T} + V_{D,p}\frac{C_D}{C_T} + V_{CG,p}\frac{C_{CG}}{C_T} \quad (7)$$

where $V_{Tun,p}$, $V_{S,p}$, $V_{D,p}$ and $V_{CG,p}$ are the voltages applied to the tunneling, source, drain, and control gate terminals, respectively, during the programming/erase phase. The initial charge $Q_0$, can be calculated by measuring the threshold voltage, $V_{Th,0}$, on a standalone transistor with the same geometry of $M_C$ at both 300 K and 15 K. The charge $Q_0$ is therefore given by:

$$Q_0 = V_{Th,0}C_T - V_{Th}(0)C_{CG} + V_{S,r}(C_D + C_{Tun}) - V_{D,r}C_D - V_{Tun,r}C_{Tun} \quad (8)$$

where $V_{Th}(0)$ is the initial threshold voltage of the device by controlling the transistor $M_C$ with the control gate terminal. $V_{Tun,r}$, $V_{S,r}$ and $V_{D,r}$ are the voltages applied to the tunneling, source and drain terminals, respectively, during the measurement of the transfer-characteristic curve of $M_C$.

Now we can solve the Eq. (6) in order to obtain the following temporal dependence of $V_{FG}$:

$$\int_{V_{FG}(0)}^{V_{FG}(t)} \frac{dV_{FG}}{I_{TUN/IHEI}(V_{FG})} = \frac{t}{C_T} \quad (9)$$

where t is an integer multiple of the time duration of a single programming/erase phase. The time evolution of $V_{FG}$ extracted from measurements with different operating conditions of the device, allow us to fit the Eq. (9) and find a unique set of fitting parameters for Eqs. (1) and (2).

*4. Experimental results*: Here, we show the experimental characterization of the device under test, shown in Figure 3, and the model validation of the Eqs. (1) and (2) both at 300 K and 15 K.

Figure 4a shows the transfer-characteristic curves of $M_C$ measured at different charges trapped in the floating gate. A sequence of voltage pulses of 7.2 V and duration $T_p$= 100 ms is applied to the Tun terminal. After each pulse, we measured the drain current of $M_C$ by applying a voltage sweep to the control gate terminal CG with a drain-source voltage fixed at 50 mV. From the transfer-characteristic curve of $M_C$ biased in the linear region, the threshold voltage $V_{Th}$ was extracted with the ELR method reported in [21]. The $V_{Th}(n)$ as a function of the number of pulses measured on the same device both at 300 K and 15 K is reported in Fig. 4b. At the beginning of the two experiments, the amplitude of the Tun voltage was selected to have the same voltage drop across $M_T$ by estimating the floating gate voltage $V_{FG}$ with the Eqs. (5), (7) and (8). Although the initial potential barrier q($V_{Tun}$-$V_{FG}$) driving the tunneling mechanism is the same, the tunneling process is less efficient at low temperature due to the lower thermal energy of the electrons. This experimental evidence shows the possibility to finely tune the threshold voltage of the device at a very low temperature, which is prospected to be a promising ingredient in the design of programmable readout circuit of solid state qubits.

In order to validate the compact model given in Eq. (1), we extracted the global parameters, reported in Table 1, by measuring the evolution of the threshold voltage $V_{Th}$ under different biasing conditions and different duration $T_P$ of the applied pulse.
the

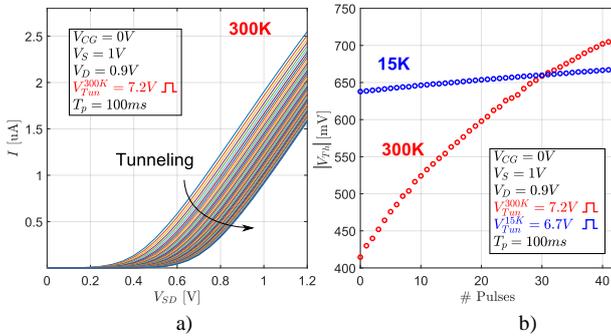

**Fig. 4** *a) Transfer-characteristic curves of $M_C$ after each tunneling pulse applied to the device biased with the reported values. b) Evolution of the threshold voltage as a function of the number of pulses at 300 K (red) and 15 K (blue).*

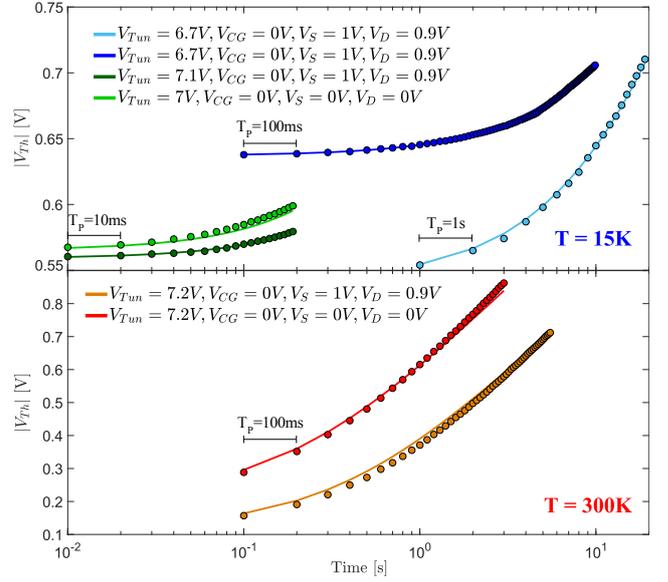

**Fig. 5** *Increasing of the threshold voltage $V_{Th}$ of $M_C$ measured (dots) and modeled (continuous lines) by Eq. (1) after each tunneling pulse, under different bias conditions and different duration $T_P$ of the tunneling pulse, both at 300 K (bottom) and 15 K (top). All measurements at a given temperature are fitted with a unique set of equation parameters.*

The starting $V_{Th}$ value is unimportant, provided the initial charge trapped into the floating gate node did not prevent the extraction of electrons. In Figure 5, the dots are the measured data and the continuous lines the fits from Eq. (1).

The low level of electrons that tunnel out of the floating gate can be appreciated in Fig. 8a, where the tunneling currents estimated from measurements are reported both at 300 K and 15 K as a function of the voltage drop across the oxide, $V_{TUN}$-$V_{FG}$. The values were estimated from eq. (6) assuming a constant current during the tunneling phase of duration $T_p$: $I_{TUN} \approx C_T \frac{\Delta V_{FG}}{T_p}$, where $\Delta V_{FG}$ is obtained by the threshold voltage difference before and after the pulse. The estimated tunneling currents for different working conditions are in good agreement with the model in eq. (1), reported as a dashed line in Fig. 6a. Fig. 5 and Fig. 6a show clearly that the model of Eq. (1) ordinarily used at room temperature is also valid at 15 K.

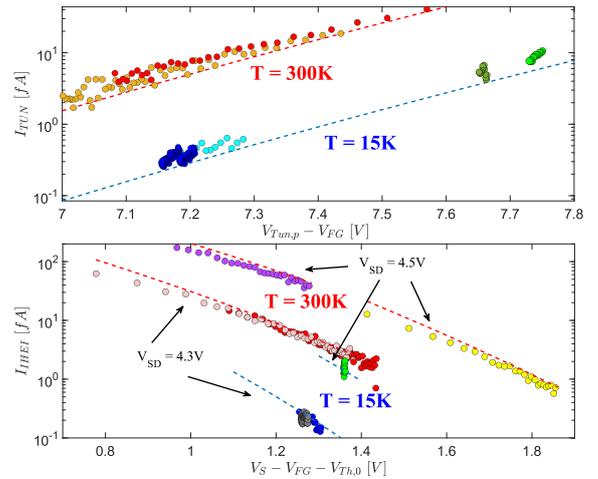

**Fig. 6** *a) Tunneling current measured (dots) and modeled (dashed lines) both at 300 K and 15 K. The color legend is the same as in Fig. 5. b) IHEI current measured (dots) and modeled (dashed lines) both at 300 K and 15 K for two different $V_{SD}$. The color legend is the same as in Fig. 8*

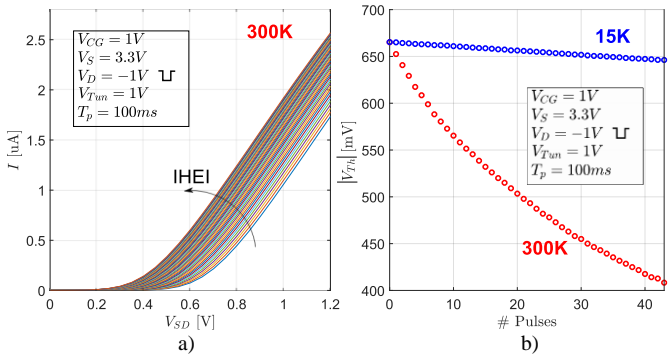

**Fig. 7** *a) Transfer-characteristic curve after each hot-electron injection pulse applied to the device. b) Threshold voltage $V_{Th}$ evolution as a function of the number of pulses at 300 K (red) and 15 K (blue). The values reported in the legend are referred to the operating condition in the erase phase.*

A similar analysis was performed for the reduction of the threshold voltage through the injection of hot electrons, which has required the measurement of the drain current of $M_C$ during the IHEI in order to fit the Eq. (2).

Figure 7a shows the measured I-$V_{SD}$ curves at 300 K after each voltage pulse applied to the drain terminal for a time duration of $T_P$=100 ms. The corresponding change of $V_{Th}(n)$ as a function of the number of pulses is reported in Figure 7b both at 300 K and 15 K, to evaluate the different efficiency of the hot electron injection . Despite the initial operating conditions of the device are the same in terms of $V_{ov}$ and $V_{SD}$, the injection current is smaller at 15 K, due to the lower thermal energy of the carriers. Different bias conditions and pulse durations were explored both at 300 K and 15 K.

Figure 8 shows the measured threshold voltage (dots) after each hot-electron injection pulse. The measurements allow the extraction of the fitting parameters of Eq. (2), obtaining two sets of parameters, one for 300 K and one for 15 K. They are reported in Table 1. The resulting fit is shown in Figure 8 by a continuous line. At 15 K, the resolution in the tunability of $V_{Th}$ reached in our measurements can be deduced from the curve with $V_D$ at -1.1 V. The minimum variation of $V_{Th}$ measured with a single pulse has a mean value of only 150µV and a standard deviation of 180µV. We envisage a further improvement of the resolution by using larger floating gates and analog techniques of compensation. In the [13], sensitivity on the signal readout of 100µV is experimentally demonstrated at 300 K, suggesting an improved programming sensitivity at lower temperatures being the FN tunneling and IHEI slower.

Figure 6b reports the values of the IHEI current at 300 K and 15 K estimated with the same assumptions of Fig. 6a. The graph confirms a decrease of about a factor 10 of the IHEI efficiency at low temperature. Fig. 6b also reports as dashed lines the values predicted by Eq. (2). The very good agreement between the experimental data and the fit validates the proposed compact model.

|  | 300 K | 15 K |
|---|---|---|
| **Tunneling** | | |
| $I_{Tun0}$ [$fA \cdot \mu m^{-2} \cdot V^{-2}$] | 3.5 | 1.05 |
| $B$ [$V \cdot cm^{-1}$] | 4.03E8 | 4.2E8 |
| **IHEI** | | |
| $C$ [$V^{-3}$] | 3.1E3 | 1.3E3 |
| $D$ [$V$] | 118.9 | 132.3 |
| $\delta$ | 0.71 | 0.85 |

**Table 1:** *fitting parameters of Eqs. (1) and (2) extracted from measurements both at 300 K and 15 K*

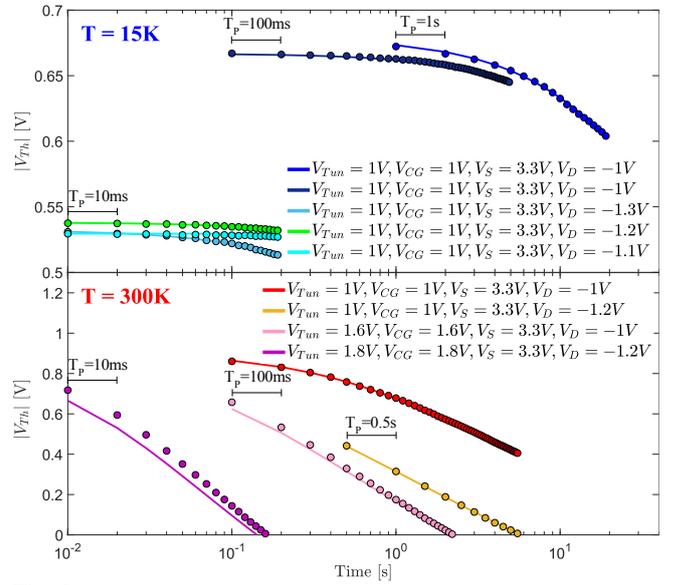

**Fig. 8** *Decrease of the threshold voltage $V_{Th}$ of $M_T$ measured (dots) and modeled (continuous lines) by Eq. (2) after each IHEI event, under different bias conditions and with different duration $T_P$ of the IHEI pulse, both at 300 K (bottom) and 15 K (top). All measurements at a given temperature are fitted with a unique set of equation parameters.*

*Conclusion*: We demonstrate that standard CMOS technology is suitable to implement a non-volatile cryogenic analog memory based on floating-gate transistors. The excellent tunability of the threshold voltage offered by IHEI and tunneling mechanism suggests the possibility to achieve a cryogenic analog memory that can be programmed or erased with a high resolution. The compact models of Eqs. (1, 2) were validated under different operation conditions, enabling a precise simulation of programmable analog circuits based on the proposed floating-gate device, even at very low temperatures. The results facilitate the design of accurate, compact, low power and configurable analog cryogenic circuits as required for the readout of silicon qubits. The results pave the way to alternative design of configurable analog circuits for an accurate readout of the silicon qubits.

*Acknowledgments:* This work was supported by QUASIX Grant from Italian Space Agency. This work was partially performed at Polifab, the micro- and nanofabrication facility of Politecnico di Milano.

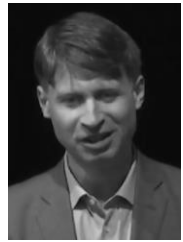

**Enrico Prati** (1974) is senior research scientist of CNR. He received the Italian Laurea Degree in Theoretical Physics in 1998 at the University of Pisa and, in 2002, the PhD in Physics. His current research fields are both theoretical and experimental aspects of silicon quantum devices and quantum artificial intelligence. In February 2004 he has been awarded with the Young Scientist Award 2004 by the URSI Committee. In March 2009 he received the 4th Jury Prize for its Essay on the Nature of Time from the FQXi granted by the John Templeton Foundation. He has been plenary keynote speaker at IEDM 2014 and TEDx speaker in 2016. He is author of about 100 journal and conference papers, 10 chapters of book and 3 patents. He is co-editor of the book Single Atom Nanoelectronics.

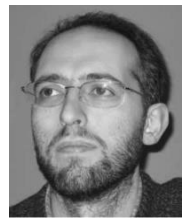

**Giorgio Ferrari** (1973) received the Laurea degree and the Ph.D. degree from Politecnico di Milano in electronics engineering in 1999 and 2003, respectively. Since 2005, he has been an Assistant Professor of electronics at Politecnico di Milano. His research interests are in the areas of development of very low noise instrumentation for high sensitivity electrical measurements both at room temperature and cryogenic temperature. He has designed and operated an innovative spectrum analyzer based on correlation techniques. He has contributed to implement a new current-sensing atomic force microscope able to perform quantitative impedance and capacitance measurements on nanometric areas. He is co-author of more than 150 journal and conference papers, of 8 chapter books and holds 9 patents.

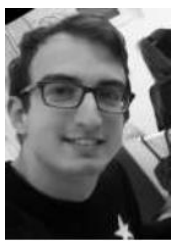

**Michele Castriotta** (1995) is a Ph.D student in Information Technology at the Department of Electronics, Information and Bioengineering (DEIB) of Politecnico di Milano, Italy. He received the M.Sc. (2020) Degrees in Electronics Engineering from Politecnico di Milano. His current research field is related to the development of very low noise instrumentation for high sensitivity electrical measurements at cryogenic temperature. He also investigates the aspects of the quantum transport in nano-scale devices.